\documentclass[12pt,aps,nofootinbib,preprintnumbers,floatfix]{revtex4}
\usepackage{amsmath,amssymb,graphicx,url}
\begin{document}
%%%%%%%%%%%%%%%%%%%%%%%%%%%%%%%%%%%%%%%%
\title{On the energy deposited by a quark moving in an $\mathcal{N}=4$ SYM plasma }
\author{Amos Yarom}
\affiliation{Ludwig-Maximilians-Universit\"at \\ Department f\"ur Physik, \\ Theresienstrasse 37, 80333 M\"unchen, Germany}
\email{yarom@theorie.physik.uni-muenchen.de}

\preprint{LMU-ASC 16/07}

\begin{abstract}
We evaluate the energy momentum tensor of a massive quark as it moves through an $\mathcal{N}=4$ SYM quark gluon plasma at constant velocity. We find that in the near-quark region, where the dynamics is expected to be dominated by dissipative behavior, the energy density may be quantitatively characterized by a transient at velocities above the speed of sound of the plasma.
\end{abstract}

%%%%%%%%%%%%%%%%%%%%%%%%%%%%%%%%%%%%%%%%%

\maketitle

%%%%%%%%%%%%%%%%%%%%%%%%%%%%%%%%%%%%%%
%%%%%%%%%%%%%%%%%%%%%%%%%%%%%%%%%%%%%%
\section{Introduction and summary}
%%%%%%%%%%%%%%%%%%%%%%%%%%%%%%%%%%%%%%
%%%%%%%%%%%%%%%%%%%%%%%%%%%%%%%%%%%%%%
The Relativistic Heavy Ion Collider (RHIC) at the Brookhaven national laboratory \cite{BRAHMS,PHENIX,PHOBOS,STAR} (see \cite{RHICreview} for a review) probes the dynamics of an expanding quark gluon plasma (QGP) at temperatures slightly above the QCD deconfinement temperature $T_{QCD} \sim 170$~MeV. One of the prominent features of the QGP at RHIC is the quenching (lack of energy) of hadronic jets emitted from the plasma. It is understood that the plasma acts as a dissipative medium for the partons moving through it. The exact mechanism for the energy loss of the partons is unclear partly because the coupling constant $g_{YM}$ is quite large and certain features of the QGP are not accessible perturbatively, and partly due to the difficulty in modeling the interaction of a parton with the surrounding quark gluon plasma.

One approach which overcomes these difficulties is to consider, instead, a quark in a strongly coupled $\mathcal{N}=4$ large $N$ $SU(N)$ SYM plasma. While there are several differences between the former and a QCD plasma, one of the lessons that has been learned over the past year is that many features of an $\mathcal{N}=4$ SYM plasma roughly coincide with those of the QGP at RHIC. A comparison includes the energy density, the viscosity to entropy density ratio \cite{etatos1,etatos2}, the friction coefficient for the moving quark \cite{Washington,Gubserdrag} and the jet quenching parameter \cite{jq1,other4}. One may extend the analysis to include theories which break conformal invariance or $\mathcal{N}=4$ supersymmetry. For example \cite{etatos3} and \cite{etatos4} study the viscosity to entropy density ratio in a non conformal theory and a theory with massive quarks in the fundamental representation. The diffusion coefficient for various deformations of the superconformal $\mathcal{N}=4$ theory have been calculated and analyzed in \cite{diff1,drag1,drag2,drag3,drag4,drag5,drag6,drag7,drag8}. Similar investigations of the jet quenching parameter have been carried out in \cite{jq2,jq3,jq4,jq5,jq6,jq7,jq8,jq9}. Some other features of the quark gluon plasma in $\mathcal{N}=4$ SYM and related theories have been studied in \cite{meson1,sl1,sl2,other1,other2,other3,Plasmaballs,Gubserdilaton,Gubseremtensor,cpotentialdilaton,mydilaton}.

This is not to say that $\mathcal{N}=4$ SYM theory is a good approximation to QCD. While there is qualitative agreement between the two, on the quantitative level there is the expected mismatch (see \cite{other4} and \cite{Gubsercomparison} for some discussions). Here, we shall adopt a somewhat conservative approach and try to understand the characteristics of the energy deposition of a moving quark in an $\mathcal{N}=4$ SYM plasma as a first step towards understanding quark-QGP dynamics. While being motivated by the comparisons between $\mathcal{N}=4$ SYM and the results of the RHIC experiment listed above, whether our analysis is actually related to QCD or not remains somewhat speculative.

To study the energy associated with a quark we shall evaluate the near-quark energy momentum tensor as the quark travels through a strongly coupled QGP at constant velocity. This may be done by appealing to the AdS/CFT duality \cite{AdSCFT}. There, one may evaluate the energy momentum tensor of a heavy quark in $\mathcal{N}=4$ SYM at finite temperature by considering metric fluctuations due to a string hanging down from the boundary of an AdS-Schwarzschild (AdS-SS) background geometry \cite{BalKra,Maldacenawilson,Gubseremtensor}.

We are interested in the temperature dependent contribution of the quark motion to the energy momentum tensor. Apart from this subleading term, the energy momentum tensor will also be influenced by the leading zero temperature contribution of the moving quark,\footnote{The value of the leading zero temperature contribution of the moving quark to the energy momentum tensor may be obtained up to an overall constant by imposing conformal and Poincar\`e symmetry \cite{Gubseremtensor}. We reproduce this result in equation (\ref{E:emtensor}).} and by the QGP itself. Our main result is the real space value of the energy momentum tensor $\langle T_{\mu\nu}\rangle\Big|_{d}$ (the `d' is a reminder that the energy momentum tensor of the conformal plasma and the leading contribution from the quark motion at zero temperature have been subtracted) which is valid in the vicinity of the quark \begin{subequations}
\label{E:main1}
\begin{align}
	\frac{\sqrt{1-v^2}}{\sqrt{\lambda} T^2}\langle T_{tt} \rangle\Big|_{d}& = \frac{v  \left(r^2(-5+13v^2-8v^4)+(-5+11v^2)x_-^2\right)x_-}{72\left(r^2(1-v^2)+x_-^2\right)^{5/2}},\\
	\frac{\sqrt{1-v^2}}{\sqrt{\lambda} T^2}\langle T_{t x_-} \rangle\Big|_{d}& = -\frac{v^2 \left(2 x_-^2+(1-v^2)r^2\right)x_-}{24 \left(r^2(1-v^2)+x_-^2\right)^{5/2}},\\	
	\frac{\sqrt{1-v^2}}{\sqrt{\lambda} T^2}\langle T_{tr} \rangle\Big|_{d}& = -\frac{(1-v^2) v^2 \left(11 x_-^2+8r^2(1-v^2)\right) r}{72 \left(r^2(1-v^2)+x_-^2\right)^{5/2}},\\	
	\frac{\sqrt{1-v^2}}{\sqrt{\lambda} T^2}\langle T_{x_- x_-} \rangle\Big|_{d}& = \frac{v  \left(r^2(8-13 v^2 +5v^4)+(11-5v^2)x_-^2\right)x_-}{72 \left(r^2(1-v^2)+x_-^2\right)^{5/2}},\\
	\frac{\sqrt{1-v^2}}{\sqrt{\lambda} T^2}\langle T_{x_- r} \rangle\Big|_{d}& = \frac{ v (1-v^2) \left(8 r^2(1-v^2)+11 x_-^2\right)r}{72 \left(r^2(1-v^2)+x_-^2\right)^{5/2}},\\
	\frac{\sqrt{1-v^2}}{\sqrt{\lambda} T^2}\langle T_{r r} \rangle\Big|_{d}& = -\frac{v (1-v^2)  \left(5 r^2(1-v^2) + 8 x_-^2\right)x_-}{72 \left(r^2(1-v^2)+x_-^2\right)^{5/2}},\\
	\frac{\sqrt{1-v^2}}{\sqrt{\lambda} T^2}\langle T_{\theta \theta} \rangle\Big|_{d}& = -\frac{v (1-v^2) x_- }{9 \left(r^2(1-v^2)+x_-^2\right)^{3/2}}.
\end{align}
\end{subequations}
In equations (\ref{E:main1}) $\mathcal{O}(T^4)$ corrections to $\langle T_{\mu\nu}\rangle\Big|_{d}$ have been omitted and contributions with delta-function support where ignored.

%Other formats
%\begin{multline}
%	\frac{72 \rho^5 \sqrt{1-v^2}}{\sqrt{\lambda}T^2} \langle T_{\mu\nu} \rangle \Big|_{d}
%	\\=
%	\begin{pmatrix}
%		v x_- \left(r^2(-5+13v^2-8v^4)+(-5+11v^2)x_-^2\right) &
%		-3 v^2 x_-\left(2 x_-^2+(1-v^2)r^2\right) &	
%		-(1-v^2) r v^2 \left(11 x_-^2+8r^2(1-v^2)\right) &
%		0
%	\\
%		-3 v^2 x_-\left(2 x_-^2+(1-v^2)r^2\right) &	
%		v x_- \left(r^2(8-13 v^2 +5v^4)+(11-5v^2)x_-^2\right) &
%		r v (1-v^2) \left(8 r^2(1-v^2)+11 x_-^2\right) &
%		0
%	\\
%		-(1-v^2) r v^2 \left(11 x_-^2+8r^2(1-v^2)\right) &
%		r v (1-v^2) \left(8 r^2(1-v^2)+11 x_-^2\right) &
%		-v (1-v^2) x_- \left(5 r^2(1-v^2) + 8 x_1^2\right)&
%		0
%	\\
%		0 & 0 & 0 &
%		- 8 \rho^2 r^2 x_- v (1-v^2)
%	\end{pmatrix}
%\end{multline}

Our notation is such that $x_- = x-vt$ is the comoving coordinate of the quark, $T$ is the temperature of the quark gluon plasma, $v$ the velocity of the quark (in the $x$ direction), $r$ is the direction transverse to the quark motion, $\theta$ is the polar angle and $\lambda$ is the large $N$ 't Hooft coupling constant.

An analysis of $\langle T_{\mu\nu}\rangle$ was initiated in \cite{Gubseremtensor} where the large distance behavior of the energy momentum tensor was discussed. There it was shown that if the quark moves faster than the speed of sound, $v^2=1/3$, a Mach cone will be created far from the moving quark. This fits nicely with a hydrodynamic model \cite{sound1} (valid at large distances as well) for the sound waves carried by the plasma as a result of the quark motion. As argued in \cite{sound1} when one gets closer to the moving quark he or she will reach a regime where hydrodynamics becomes non-linear and turbulent effects are expected. In the immediate vicinity of the quark there is a region where the hydrodynamic approximation breaks down completely and strong dissipative effects dominate the dynamics. This is precisely the region described by equations (\ref{E:main1}).
It is nice to see that the AdS/CFT duality provides for a method of probing
the near-quark surroundings which is not accessible via standard constructions due to strong dissipative behavior (see \cite{sound1}) and the lack of a model for the relevant interactions.

\begin{figure}[hbt]
\begin{center}
%\scalebox{0.48}{\includegraphics{edeposit.eps}}
\scalebox{0.48}{\includegraphics{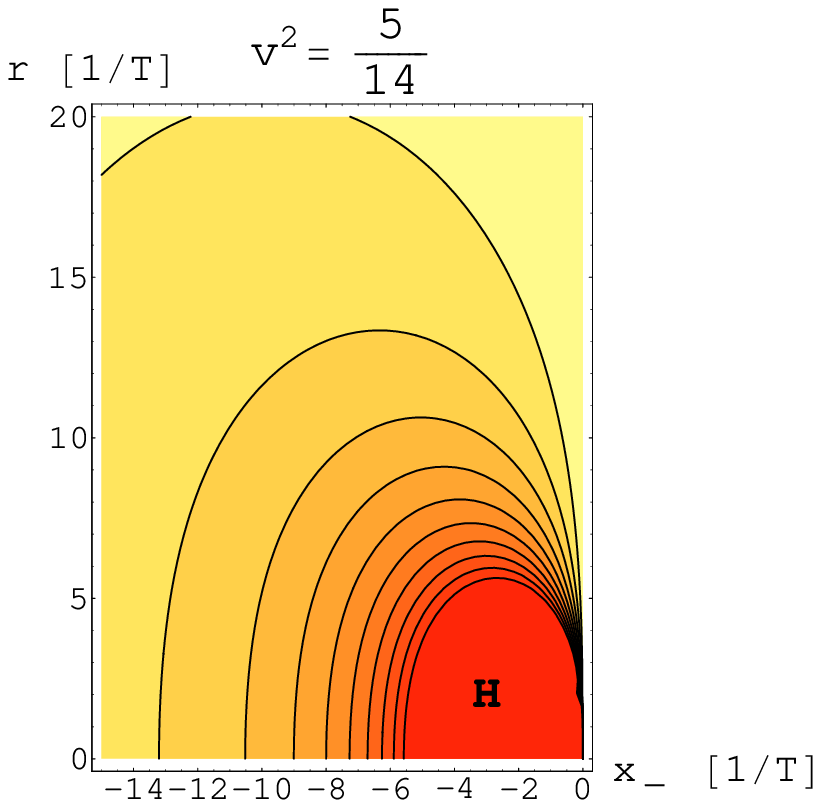}}\scalebox{0.48}{\includegraphics{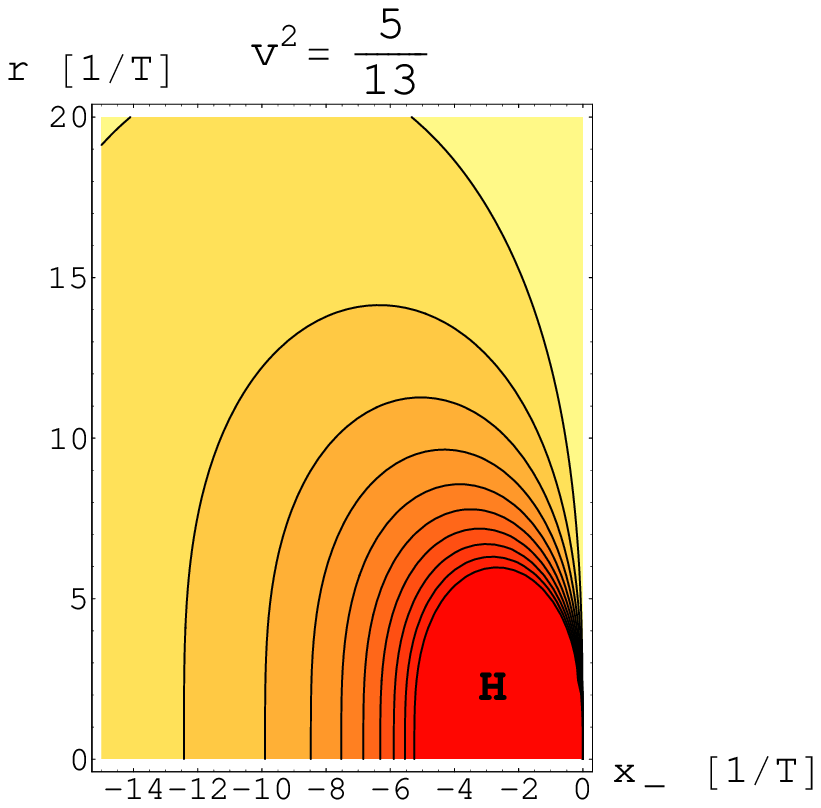}}
\scalebox{0.48}{\includegraphics{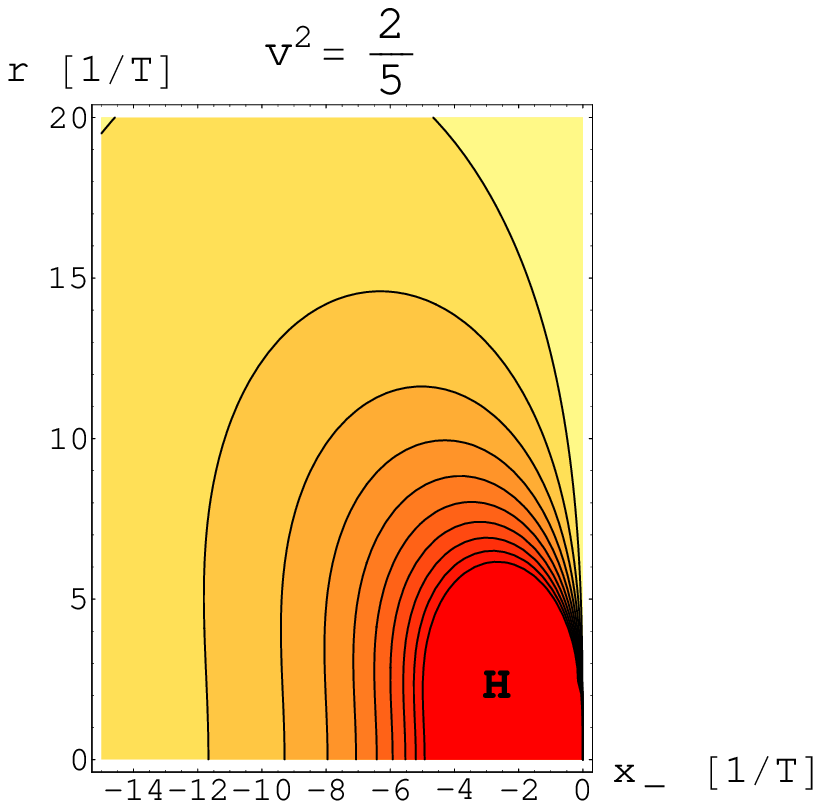}}\scalebox{0.48}{\includegraphics{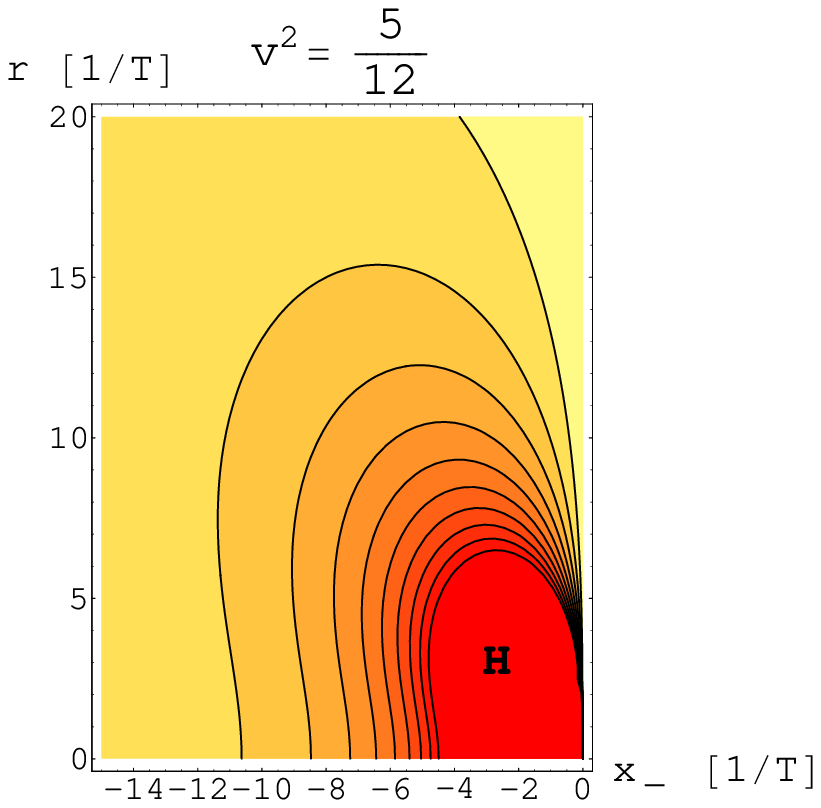}}\\
\scalebox{0.48}{\includegraphics{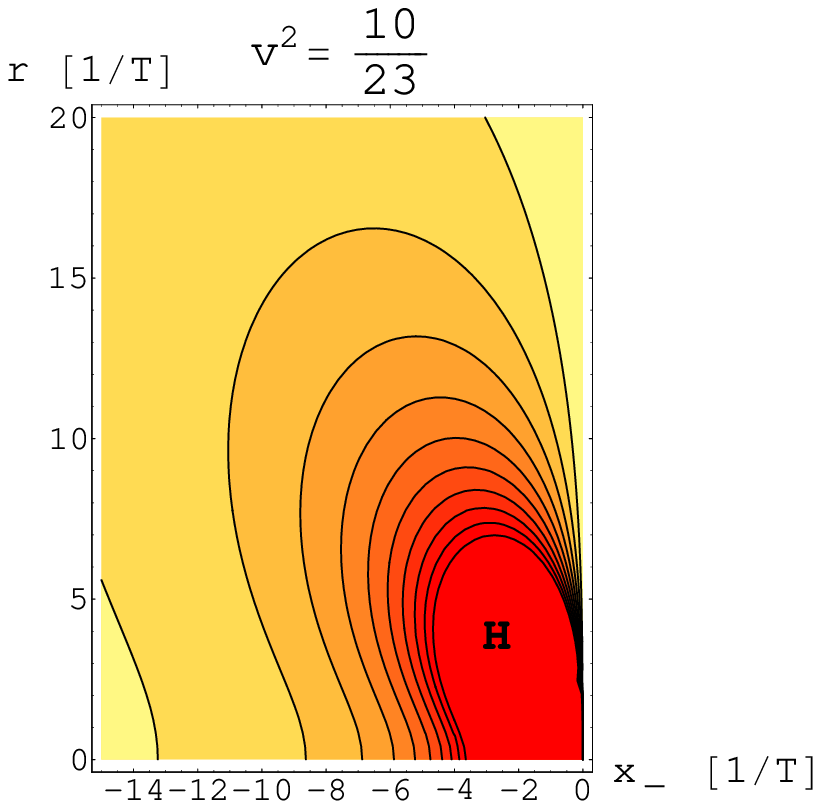}}\scalebox{0.48}{\includegraphics{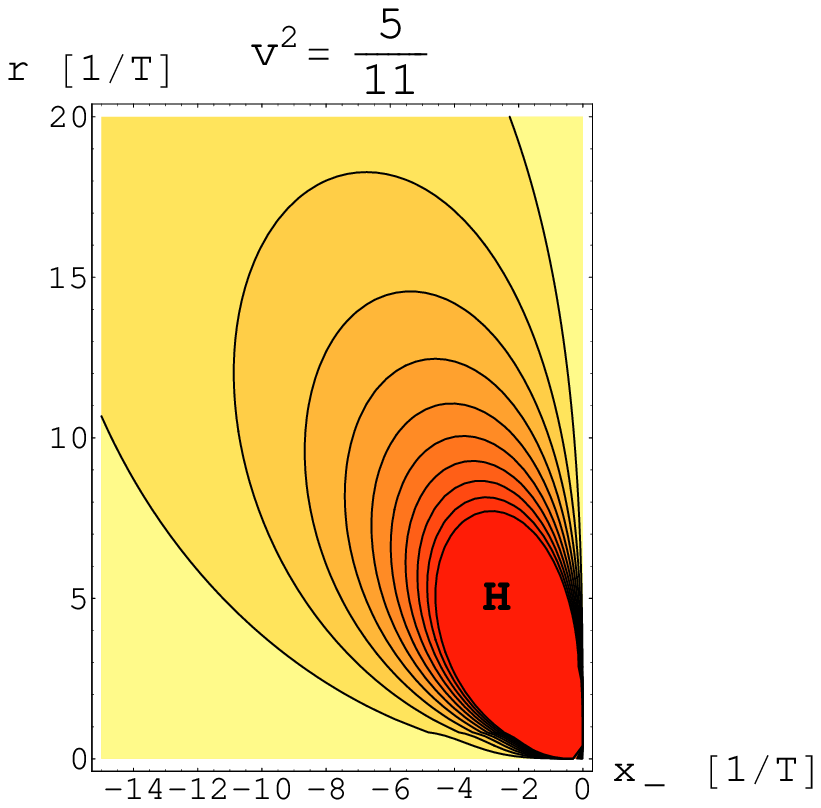}}
\scalebox{0.48}{\includegraphics{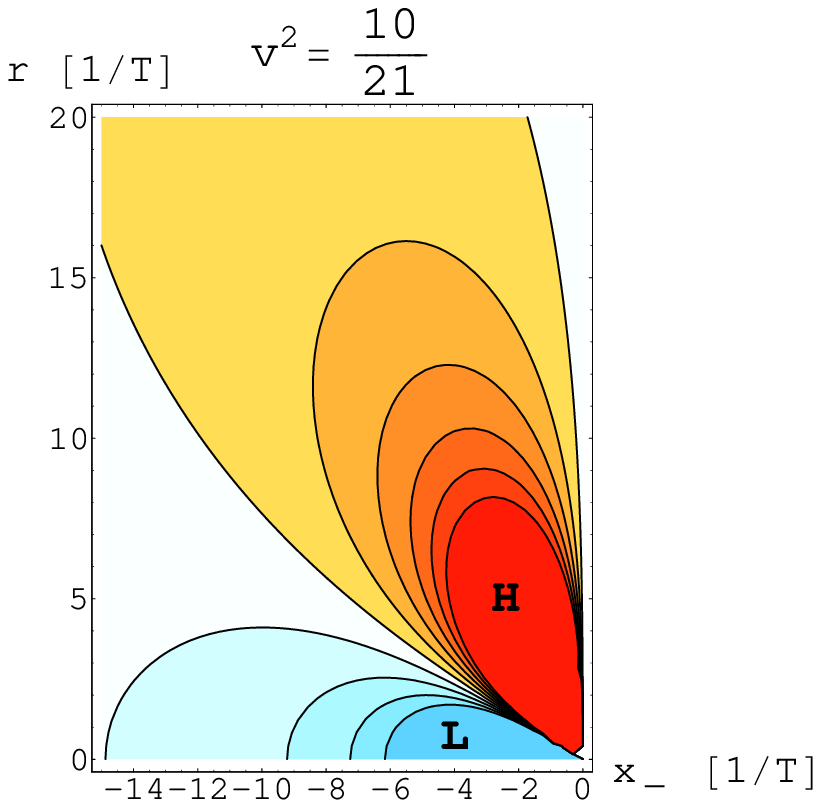}}\scalebox{0.48}{\includegraphics{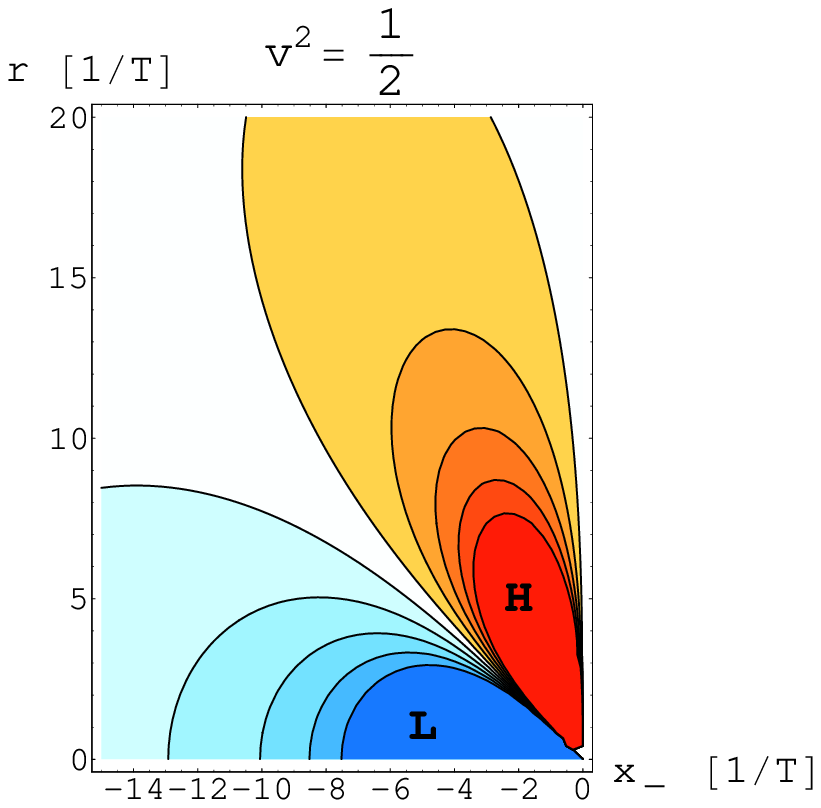}}\\
\scalebox{0.48}{\includegraphics{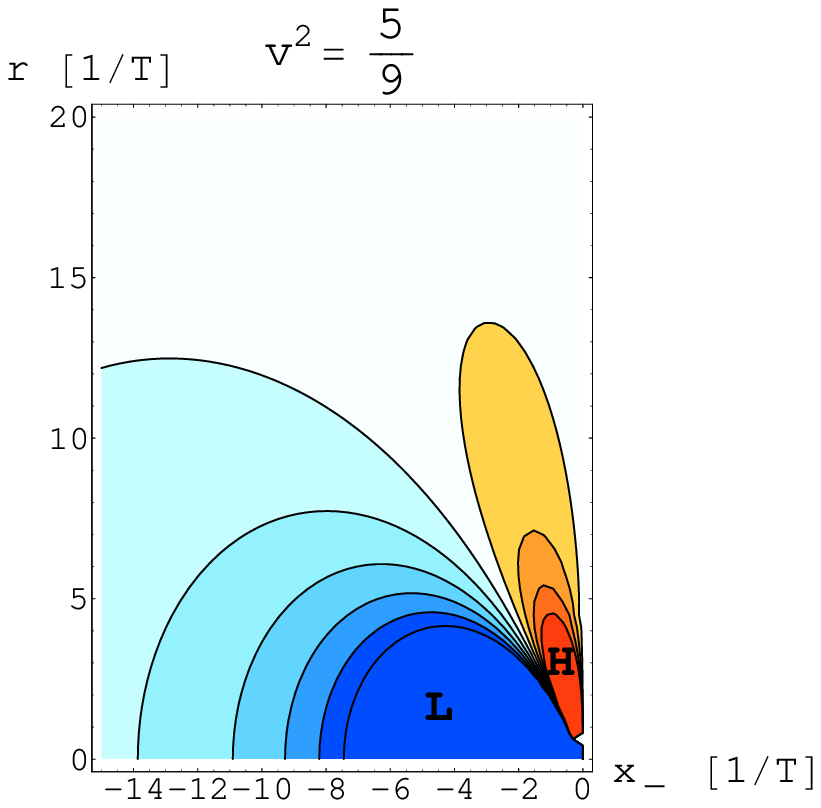}}\scalebox{0.48}{\includegraphics{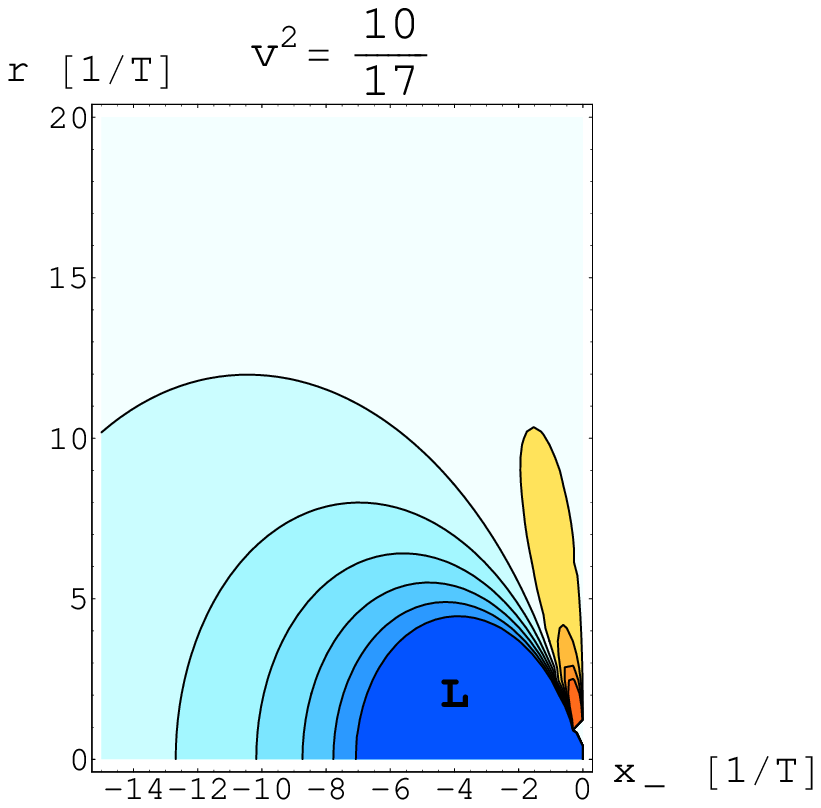}}
\scalebox{0.48}{\includegraphics{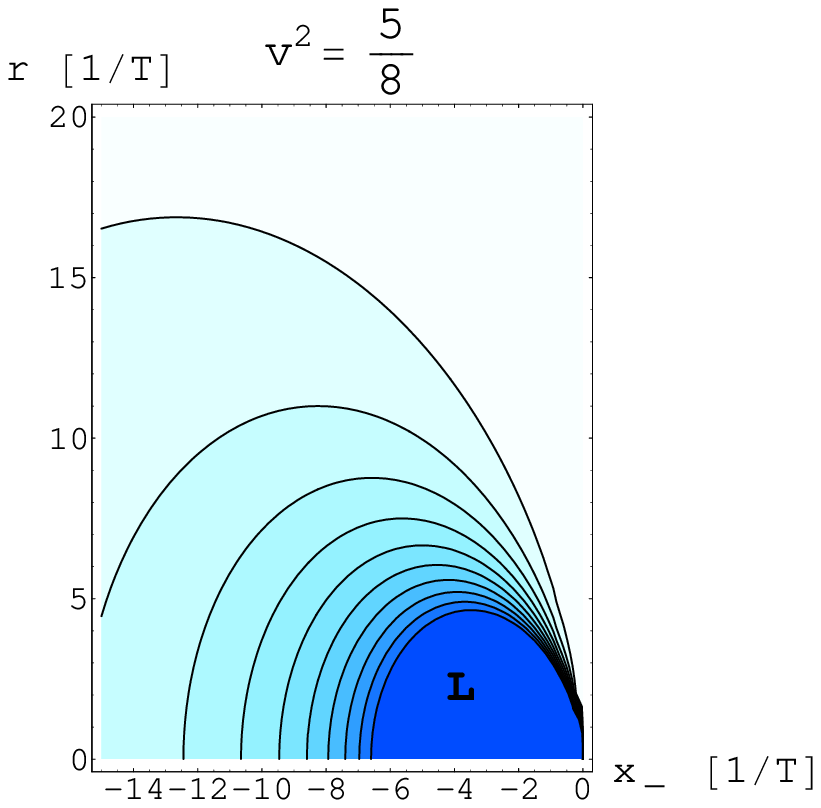}}\scalebox{0.48}{\includegraphics{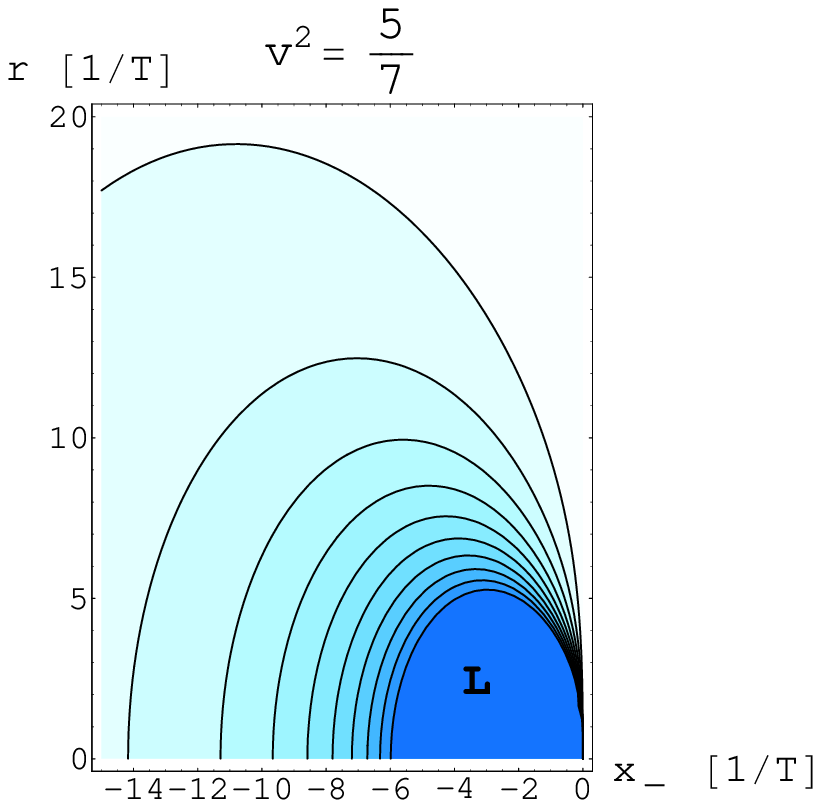}}
\caption{\label{F:plots} Contour plots of the leading dissipative contribution to the energy density
$\frac{1}{\sqrt{\lambda}T^4}\langle T_{tt}(x_-,r) \rangle \Big|_{d}$ due to the motion of a quark moving at constant velocity $v$. The horizontal axis
corresponds to the comoving coordinate of the quark, $x_- = x-vt$, while the vertical one corresponds to the direction transverse to the quark motion, $r$. Large positive values are shaded in red and marked with an `H' while lower values correspond to blue shading and are marked `L'. See \cite{emanimation} for an animation.}
\end{center}
\end{figure}

Focusing on the energy density $\langle T_{tt}(x_-,r)\rangle\Big|_{d}$ we find that it exhibits some velocity dependent directional features (see figure \ref{F:plots}) which do not appear for the other components of the energy momentum tensor. As long as the velocity is relatively low, $0<v^2<5/13$, the energy density has a vanishing gradient in the $x_-$ direction along
\[
	\tan^2 \omega_1 = \left(\frac{r}{x_-}\right)^2 = \frac{5+v^2-\sqrt{15}\sqrt{15-50v^2+47v^4}}{-20+44 v^2}(1-v^2).
\]
In this range of velocities no exceptional behavior is observed even when the quark passes the speed of sound $v^2=1/3$. As mentioned earlier, this should be contrasted with the Mach cone which is formed far from the quark as it passes the speed of sound of the conformal plasma \cite{Gubseremtensor,sound1}.

At velocities squared above $5/13$, the subleading dissipative contribution to the energy density begins developing lobe-like features specified by a non trivial, vanishing gradient in the $r$ direction at an angle of
\[
	\tan^2 \omega_2 = \left(\frac{r}{x_-}\right)^2 = \frac{5-13 v^2+8 v^4}{-5+13 v^2}.
\]

As the velocity squared of the quark is increased to $5/11$ one finds that there are two contributions to $\langle T_{tt} \rangle\Big|_{d}$, the previous contribution now deformed to a lobe of high energy, and a new low energy contribution right behind the quark.\footnote{The terms high and low energy are, of course, relative to the leading zero temperature contribution of the quark and the contribution of the plasma which have been subtracted. We would like to thank the authors of \cite{Gubsermasterequations} for clearing up a sign mixup in this context.} This new contribution is directed at an angle of
\[
	\tan^2 \omega_3 = \left(\frac{r}{x_-}\right)^2 =\frac{5+v^2+\sqrt{15}\sqrt{15-50v^2+47v^4}}{-20+44 v^2}(1-v^2).
\]

As the velocity is increased further, the  energy deficit behind the quark dominates and the contribution of the lobe becomes smaller until it vanishes at a velocity squared of $5/8$. We encourage the reader to have a look at \cite{emanimation} where this effect can be observed as a continuous function of the velocity.

This behavior may suggest that there are two mechanisms that come into play. One at relatively low velocities $v^2<5/13$, responsible for a region of high energy, and one at high velocities $v^2>5/8$, responsible for a region of energy depletion, with some overlap. It is interesting to note that the overlap region is above the Mach velocity for the quark, and that there is no remnant of the Mach cone discussed in \cite{Gubseremtensor}.\footnote{Note however, that when the Mach cone is generated there is a buildup of energy in its vicinity (see figure 2, and equation (80) of \cite{Gubseremtensor}) which may extend all the way down to the near quark region at the intermediate velocity range $5/11<v^2<5/8$.} A different possibility is that in figure \ref{F:plots} we are seeing an interference pattern of some sort. Though then it is not clear what sort of waves are propagating in the plasma since sound waves are not expected at such high momentum scales and the gluon field strength does not exhibit lobe-like behavior at short distances \cite{mydilaton}. It is perhaps worth mentioning that a region of energy deficiency behind the quark is characteristic of an LPM (Landau-Pomeranchuk-Migdal) effect \cite{lpm1,lpm2} where the parton motion in the plasma is modeled by inelastic collisions with fixed scatterers. The literature contains many more interesting discussions regarding the mechanism for energy loss in a QGP (see for example \cite{sound1,brems1,brems2,cher1}) though we could not find quantitative correlations between the descriptions there and the behavior described above.

While contemplating over these issues, the authors of \cite{Gubsermasterequations} brought to my attention that their work has some overlap with this paper. I thank them for sharing their results with me prior to publication.

%%%%%%%%%%%%%%%%%%%%%%%%%%%%%%%%%%%%%
%%%%%%%%%%%%%%%%%%%%%%%%%%%%%%%%%%%%%%
\section{\label{S:metricfluctuations} Evaluation of $\langle T_{\mu\nu} \rangle$.}
%%%%%%%%%%%%%%%%%%%%%%%%%%%%%%%%%%%%%%
%%%%%%%%%%%%%%%%%%%%%%%%%%%%%%%%%%%%%%
We wish to evaluate the expectation value of the energy momentum tensor in response to the motion of a quark in a thermal plasma in $\mathcal{N}=4$ SYM. Using the AdS/CFT dictionary this is given by the response of the AdS-SS metric to the motion of a string with one endpoint located on the AdS boundary, tracing the worldline of the quark.

Ignoring for the moment the effect of the string on the spacetime geometry, one can find the string's profile by minimizing the Nambu Goto action
\[
	S_{NG} = -\frac{1}{2\pi\alpha^{\prime}} \int \sqrt{-w} d\tau d\sigma
\]
in an AdS-SS background.
Hence, $w$ is the induced metric $w_{\alpha\beta} = G^{(0)}_{\mu\nu}\partial_{\alpha}X^{\mu}\partial_{\beta}X^{\nu}$ with $G^{(0)}$ the metric of an AdS-SS black hole;
$G^{(0)}$ may be read off of the line element
\[
	ds^2 = \frac{L^2}{z^2}\left(-g(z) dt^2 + dx^2 + \sum_{i=2,3} dx_i^2 + \frac{dz^2}{g(z)}\right)
\]
with $g(z)=1-\left(\frac{z}{z_0}\right)^4$. The coordinate $z$ runs from $0$ at the asymptotically AdS boundary to $z_0$ at the black hole horizon. We have set the value of the dilaton to its trivial background value since it decouples from the equation of motion at the linearized order that we are considering.

The equations of motion for the embedding functions of the string $X^{\mu}(\sigma,\tau)$ are supplemented by boundary conditions which require that the string endpoint moves at constant velocity $v$, and that no energy is fed to the string from the horizon. Using a static gauge and an ansatz $X^{\mu}(\sigma,\tau) = (\tau,\xi(\sigma)+vt,0,0,\sigma)$ one finds \cite{Gubserdrag,Washington}
\begin{equation}
\label{E:xi}
	\xi(z)=\frac{v z_0}{4}\left(\ln\frac{1-\frac{z}{z_0}}{1+\frac{z}{z_0}}+2\arctan\frac{z}{z_0}\right).
\end{equation}

The differential equation describing the combined dynamics of the string and the target space metric (within the supergravity approximation) can be found by minimizing the action
\begin{equation}
\label{E:combinedS}
	S=S_{EH}+S_{NG}
\end{equation}
where
\begin{equation}
	S_{EH}=\frac{1}{16 \pi G_5} \int \sqrt{-G}\left(R+\frac{12}{L^2}\right)d^5x.
\end{equation}
Since we will be looking for the expectation value of the energy momentum tensor, it will be enough to consider the linearized equations of motion for the metric fluctuations $h$, defined through $G=G^{(0)}+h$, in the presence of the string whose profile is determined by equation (\ref{E:xi}).

The linearized equations for $h$ resulting from minimizing (\ref{E:combinedS}) may be written in condensed form as
\begin{equation}
\label{E:EOMmetric}
	\mathcal{D}^{\mu\nu\rho\sigma}h_{\rho\sigma} = T^{\mu\nu}
\end{equation}
with $\mathcal{D}^{\mu\nu\rho\sigma}$ a second order linear differential operator. The boundary conditions for the metric fluctuations are that they vanish at the AdS boundary (implying that there is no source term for the metric) and that there are no outgoing modes coming from the horizon.
Once a solution, $h$, to equation (\ref{E:EOMmetric}) is obtained the energy momentum tensor in the SYM theory $\langle T \rangle = \langle T_{plasma} \rangle + \langle T_{quark} \rangle$ may be determined through \cite{Gubseremtensor}
\begin{equation}
\label{E:holographictmn}
	\langle T_{quark} \rangle = \frac{1}{\pi}\sqrt{\frac{\lambda}{1-v^2}} Q,
\end{equation}
where $Q_{\mu\nu}$ is obtained from $h=\ldots + Q z^4 + \ldots$ through a small $z$ expansion. $\langle T_{plasma} \rangle = \frac{\pi^2}{8}N^2 T^4\,diagonal(3,1,1,1)+\mathcal{O}(N^0)$ is the contribution of the thermal plasma to the energy momentum tensor \cite{TdofAdSCFT}. Since we are interested in the dissipative effects of the quark motion, we subtract this background value and so, for our purposes it may be ignored.

In equation (\ref{E:holographictmn}) one should use the relations
\begin{equation}
\label{E:AdSCFT}
	N^2 = \frac{\pi L^3}{2 G_5},\quad
	\frac{1}{\pi T} = z_0\quad\text{and}\quad
	\sqrt{\lambda} = \frac{L^2}{\alpha^{\prime}}.
\end{equation}
Apart from that, one should also make sure that all lower order terms in the expansion of $h$ in small values of $z$ correspond to contact terms (and therefore may be renormalized away\footnote{A holographic renormalization prescription was presented in \cite{Holren1,Holren2}, see also \cite{ABY1} for some recent applications. If one is careful when taking the limit where the AdS radial coordinate reaches the asymptotically AdS boundary then a subtraction of divergent contact terms will give the same results. See for example \cite{Gubseretal,Freedman3point,KlebanovWitten2,ABY2}.}).

In \cite{Gubseremtensor} it was shown that by choosing a gauge such that $h_{5\mu}=0$ and working in cylindrical coordinates,
the coupled equations of motion for the various components of the metric tensor decouple into three sets of second order equations and two sets of first order constraints. Explicitly, we Fourier transform the metric fluctuations,
\[
	\hat{h}(k,z) = \int h_{\mu\nu}(x,z) e^{-i(k_- x_-+k_2 x_2+k_3 x_3)}d^3k,
\]
and define
\begin{equation}
\label{E:smallbigh}
	\hat{h}_{\mu\nu}(k_-,k_{\bot},z) = \frac{4 G_5 L}{\alpha^{\prime} \sqrt{1-v^2}}\frac{1}{z^2}H_{\mu\nu}(k_-,k_{\bot},z),
\end{equation}
where we have also rotated $\hat{h}$ to a cylindrical coordinate system.
By considering the combinations
\begin{align}
\label{E:defA}
	A &= \frac{1}{2v^2}\left(-H_{11}+2\frac{k_-}{k_{\bot}} H_{12} - \left(\frac{k_-}{k_{\bot}}\right)^2 H_{22}
		+ \left(\frac{k}{k_{\bot}}\right)^2 H_{33} \right),\\
\label{E:defD}	
	\vec{D} &= \begin{pmatrix}
		D_1 \\ D_2
		  \end{pmatrix}
	  	= \begin{pmatrix}
	  	 \frac{1}{2v}\left(H_{01}-\frac{k_-}{k_{\bot}} H_{02}\right)\\
		  \frac{1}{2v^2}
		  \left(-H_{11} + \left(\frac{k_-}{k_{\bot}}-\frac{k_\bot}{k_-}\right) H_{12}
		  +H_{22} \right)
		 \end{pmatrix}
\end{align}
and
\begin{equation}
\label{E:defE}
	\vec{E} = \begin{pmatrix}
		E_1 \\ E_2 \\ E_3 \\ E_4
		\end{pmatrix}
	 = \begin{pmatrix}
	 	\frac{1}{2}\left(-\frac{3}{g} H_{00}+H_{11}+H_{22}+H_{33}\right)\\
		\frac{1}{2v}\left(H_{01}+\frac{k_{\bot}}{k_-} H_{02}\right)\\
		\frac{1}{2}\left( H_{11}+H_{22}+H_{33}\right)\\
		\frac{1}{4}\left(H_{11}\left(2-6\left(\frac{k_-}{k}\right)^2\right)
			+H_{22}\left(-4+6\left(\frac{k_-}{k}\right)^2\right)
			+2 H_{33} - 12 \frac{k_- k_{\bot}}{k^2} H_{12}\right)
	    \end{pmatrix},
\end{equation}
one finds that (\ref{E:EOMmetric}) reduces to
\begin{align}
\label{E:EOMtensor}
	\left(I\, \partial_z^2 +  \left(-\frac{3}{z}I + \frac{g^{\prime}}{g} K_X\right) \partial_z + V_X(z)\right) \vec{X} &= \frac{z}{g} e^{-i k_- \xi(z)} \vec{S}_X(z),\\
\label{E:constrainttensor}
	\left(\tilde{K}_{X}(z) \partial_z + \tilde{V}_{X}(z) \partial_z\right) \vec{X} &= z e^{-i k_- \xi(z)} \vec{\tilde{S}}_{X}(z),
\end{align}
with $\vec{X} = \vec{E},\vec{D}$ or $A$ and $g^{\prime} = \partial_z g(z)$. $I$ is the identity matrix in appropriate dimensions.
The matrices $K_X,\,V_X,\,\tilde{K}_X$ and $\tilde{V}_X$ and the vectors $S_X$ and $\vec{\tilde{S}}_X$ can be found in \cite{Gubseremtensor} and have been reproduced in appendix \ref{A:matrices} for completeness.
An important feature of equation (\ref{E:EOMtensor}) that we will use is that in the $z_0 \to \infty$ limit\footnote{One should use the dimensionless quantity $\tilde{k} z_0$ instead of $z_0$ (with $\tilde{k}$ some scale with units of inverse length) when considering an expansion at ``large'' values of $z_0$. At the moment, we shall refrain from writing this out explicitly to avoid cluttering the notation.}, $\frac{g^{\prime}}{g} K_X \to 0$.

Due to the linear relations between $\vec{E}$, $\vec{D}$, $A$ and $H$, equations (\ref{E:defA}) to (\ref{E:defE}), the fourth order contribution to a small $z$ expansion of $\hat{h}_{\mu\nu}$ can be inferred from the fourth order coefficient in an expansion of the fields $X$. Defining $X=N_X z+O_X z^2 - \frac{1}{3}P_X z^3+Q_X z^4+\mathcal{O}(z^5)$ (recall that $h(0)=0$), we find that $N_X=O_X=0$ and that the first order constraints (\ref{E:constrainttensor}) imply that
\begin{align}
\label{E:constraints1}
	P_A&=P_{D_1}=P_{D_2} = P_{E_2} = -P_{E_3} = 1,\\
	P_{E_1}&=1+v^2,\,P_{E_4}=v^2\left(3 \frac{k_-^2}{k^2}-1\right).
\end{align}
Inverting the relations between the $A$, $\vec{D}$, $\vec{E}$ variables and the $H$ variables, equations (\ref{E:defA}) to (\ref{E:defE}), one may obtain the third order term $P_{\mu\nu}$ in the small $z$ expansion of $\hat{h}$, $\hat{h}_{\mu\nu}= -\frac{1}{3}P_{\mu\nu}z^3+Q_{\mu\nu}z^4+\mathcal{O}(z^5)$,
\[
	P = \begin{pmatrix}
		4+2v^2 & -6v & 0 & 0\\
		-6v & 2+ 4v^2 & 0 & 0\\
		0 & 0 & 2-2v^2 & 0\\
		0 & 0 & 0 & 2-2v^2
	    \end{pmatrix}.
\]
Since $P$ is independent of the momenta, any divergences that it may cause will be contact terms. This allows us to use equation (\ref{E:holographictmn}).

The constraint equations (\ref{E:constrainttensor}) also relate the values of $Q_X$ to one another. One finds
\begin{subequations}
\label{E:constraints2}
\begin{align}
	Q_{D_2}&=Q_{D_1}-\frac{1}{4 i k_- v z_0^2},\\
	Q_{E_2}&=\frac{1}{2}Q_{E_1}-\frac{v}{4 i k_- z_0^2},\,Q_{E_3}=-\frac{1}{2}Q_{E_1},\\
	Q_{E_4}&=\frac{1}{2}\left(-1+3v^2\frac{k_-^2}{k^2}\right)Q_{E_1}+\frac{3 i k_- v (1+v^2)}{4 k^2 z_0^2}.
\end{align}
\end{subequations}
Therefore, we need only find $Q_{E_1}$, $Q_{D_1}$ and $Q_A$ to obtain all the fourth order terms in an expansion of $X$. Then, as discussed in the previous paragraph, one may invert equations (\ref{E:defA}) to (\ref{E:defE}) to obtain the fourth order term $Q_{\mu\nu}$ in an expansion of $\hat{h}_{\mu\nu}$. Plugging it into equation (\ref{E:holographictmn}) we will obtain the Fourier transformed energy momentum tensor we are looking for. To make our notation slightly more precise, we should have used $\hat{Q}_{\mu\nu}$ for the fourth order term in the expansion of $\hat{h}$ instead of $Q_{\mu\nu}$ which was defined below equation (\ref{E:holographictmn}) to be its real space counterpart. However, since in the rest of this section we will be working in Fourier space, we choose an unhatted notation which is less cumbersome.

To proceed, we note that the equations of motion (\ref{E:EOMtensor}) have the property that in the $z_0 \to \infty$ limit, the kinetic terms become diagonal and correspond to the kinetic terms obtained for a scalar field in an AdS background. This implies that in the $z_0 \to \infty$ limit it is more useful to change to a basis where the potential is diagonal as well. Our approach to solve (\ref{E:EOMtensor}) and (\ref{E:constrainttensor}) is to first redefine the variables $\vec{D}$ and $\vec{E}$ so that in the $z_0 \to \infty$ limit we shall get the AdS equations of motion. To do this, we multiply the matrix equation (\ref{E:EOMtensor}) for $X=\vec{D}$ from the left by
\begin{equation}
\label{E:diagD}
	d_D=\begin{pmatrix}
		\sqrt{2} \left(\alpha ^2+1\right) & -\sqrt{2} \alpha ^2 \\
		-\left(\alpha ^2+1\right) \sqrt{\frac{\alpha ^4}{\left(\alpha ^2+1\right)^2}+1}\quad &
			\left(\alpha ^2+1\right) \sqrt{\frac{\alpha ^4}{\left(\alpha ^2+1\right)^2}+1}
	\end{pmatrix}
\end{equation}
and for $X=\vec{E}$ by
\begin{equation}
\label{E:diagE}
	d_{E}=\begin{pmatrix}
	 	\frac{2}{9} \left(\alpha ^2+1\right) \left(2 \alpha ^2-1\right) &
			-\frac{4}{3} \alpha ^2 \left(2 \alpha ^2-1\right) &
			-\frac{4}{9} \left(\alpha ^2+1\right) \left(2 \alpha ^2-1\right) &
			\frac{1}{9} \left(1-2\alpha ^2\right)^2 \\
		0 & 1 & 1 & 0 \\
		-\frac{2}{3} \left(\alpha ^2+1\right) &
			\frac{4 \alpha ^4+4 \alpha ^2+3}{\alpha ^2+1} &
			\frac{4}{3} \left(\alpha ^2+1\right) &
			\frac{1}{3} \left(1-2 \alpha ^2\right) \\
		\frac{2}{3} \left(\alpha ^2+1\right) &
		-4 \alpha ^2 &
		\frac{2}{3}\left(1-2 \alpha ^2\right) &
		\frac{2}{3} \left(\alpha ^2+1\right)
	\end{pmatrix}.
\end{equation}
In equations (\ref{E:diagD}) and (\ref{E:diagE}) we have defined $\alpha = v\frac{k_{-}}{\tilde{k}}$ with $\tilde{k}^2 = k_{\bot}^2 +(1-v^2)k_-^2$.

This will diagonalize the equations of motion in the $z_0 \to \infty$ or $g \to 1$, $g' \to 0$  limit. Actually the equation of motion for $\vec{E}$ can not be diagonalized. Instead we have brought it into Jordan form. The precise equations of motion one obtains are rather messy, and we shall not write them out explicitly. What is important, is that in our new basis we are assured that the ``off diagonal'' terms are, by construction, at least of order $z_0^{-4}$. Therefore, we may first solve the diagonal equations and then use perturbation theory to find corrections. These will be at least of order $z_0^{-4}$ and will be neglected. Changing to the more convenient variables $Z = \tilde{k} z$, $Z_0 = \tilde{k} z_0$, we can concentrate on the following three equations
\begin{align}
	\left(\partial_Z^2 + \left( -\frac{3}{Z} - \frac{4 Z^3}{Z_0^4 - Z^4} \right) \partial_Z
	- \frac{Z_0^4 \left(Z_0^4-Z^4 \left(\alpha ^2+1\right)\right)}{\left(Z^4-Z_0^4\right)^2}\right) A^{\prime}
	&=\frac{Z e^{-i k_- \xi(Z/\tilde{k})}}{g(Z)\tilde{k}},
\label{E:Ap}\\
	\left(\partial_Z^2  + \left(-\frac{3}{Z} + 4 \frac{Z^3 \alpha^2}{Z_0^4-Z^4}\right) \partial_Z
	-\frac{Z_0^4 \left(Z_0^4-Z^4 \left(\alpha ^2+1\right)\right)}{\left(Z^4-Z_0^4\right)^2}\right)D_a
	&=\frac{\sqrt{2}Z e^{-i k_- \xi(Z/\tilde{k})}}{g(Z)\tilde{k}},	
\label{E:Db}	
\end{align}
\begin{multline}
	\left(
		\partial_Z^2 +
		\left(
			-\frac{3}{Z} +
			\frac{4 Z^3 \left(4 \alpha^2(\alpha^2+1)+3\right)}
			     {3 \left( Z^4-Z_0^4 \right)}
		\right)\partial_Z -		
		\left(
			\frac{Z_0^4 \left(Z_0^4-Z^4 \left(\alpha^2+1\right)\right)}
				{\left(Z^4-Z_0^4\right)^2}
		\right)
	\right) E_a
\\
	=
	\frac{Z e^{-i k_- \xi(Z/\tilde{k})}}{g(Z)\tilde{k}}
	\left(
		\frac{\left(2 \alpha ^2-1\right) \left(\left(\alpha ^2+1\right) v^2-\alpha ^2+2\right)}{3 \left(\alpha ^2+1\right)}
		-
		\frac{2 v^2 Z^4 \left(\alpha ^2+1\right) \left(2 \alpha ^2-1\right)}{3 \left(Z^4-Z_0^4\right)}
	\right),
\label{E:Ed}
\end{multline}
where
\begin{align}
\label{E:DbtoD1}
	\tilde{k}^{-2} D_a &=\sqrt{2} \left(D_1 \left(\alpha ^2+1\right)-D_2 \alpha ^2\right),\\
\label{E:EdtoE1}
	\frac{\tilde{k}^{-2} E_a}{2\alpha^2-1} &= -\frac{4}{3} E_2 \alpha^2
		+\frac{1}{9} E_4 \left(2 \alpha ^2-1\right)
		+\frac{2}{9} E_1 \left(\alpha ^2+1\right)
		-\frac{4}{9} E_3 \left(\alpha ^2+1\right),
	\\
	\tilde{k}^{-2}A^{\prime}&=A.
\end{align}
The $D_a$ and $E_a$ equations are the first diagonal components of (\ref{E:EOMtensor}) after multiplying it from the left  by $d_{D}$, equation (\ref{E:diagD}), or by $d_{E}$, equation (\ref{E:diagE}), respectively.

The $A^{\prime}$ equation of motion is identical to the equation of motion for the dilaton in an AdS-SS black hole background as it is sourced by the trailing string \cite{Danielsson,mydilaton,Gubserdilaton,Gubseremtensor}. In \cite{mydilaton} the large momentum asymptotics of equation (\ref{E:Ap}) were found and the near quark value of the field strength, $\sim Tr F^2$, was evaluated. The WKB method was used to solve the homogeneous equation at large momenta; the solution to the non homogeneous equation then follows by constructing the Greens function from the homogeneous solutions.

Alternatively, to obtain the large $Z_0$ asymptotics of $A^{\prime}$ one may expand the equation of motion (\ref{E:Ap}) and the function $A^{\prime}(Z)$ in an inverse power series in $Z_0$. The unique normalizable solution to order $Z_0^{-2}$ is given by
\begin{equation}
\label{E:soldilaton}
	A^{\prime}(Z) = \frac{\pi Z}{2 \tilde{k}} \left(Z(I_2(Z)-L_0(Z))+2 L_1(Z)\right)
	 - \frac{i \alpha}{3 Z_0^2}Z^4,
\end{equation}
where $I_2$ is a modified Bessel function of the first kind and $L_i$ are modified Struve functions of order $i$.
A caveat in this approach is that the boundary conditions are imposed at $Z_0$ which is the perturbative parameter we are expanding in. To properly impose the boundary conditions on generic solutions one should resum the whole power series. Nonetheless, the solution obtained in this somewhat hand-waving manner coincides with the one obtained using the WKB method of \cite{mydilaton}. In both cases, one finds that for
\begin{equation}
\label{E:expansionap}
	A^{\prime} = \ldots + Q_{A^{\prime}} Z^4 + \ldots,
\end{equation}
$Q_{A^{\prime}}$ is given by
\begin{subequations}
\label{E:solutionnew}
\begin{equation}
\label{E:qap}
	\tilde{k} Q_{A^{\prime}} = \frac{1}{16}\pi -\frac{i \alpha}{3 Z_0^2} + \mathcal{O}(Z_0^{-4}).
\end{equation}

In order to find the $\mathcal{O}(Z^4)$ coefficients $Q_X$ in the small $Z$ expansion of $X=D_a$ and $X=E_a$ (in analogy with equation (\ref{E:expansionap})), one needs to solve the equations of motion for $D_a$ and $E_a$, (\ref{E:Db}) and (\ref{E:Ed}).
Following the somewhat heuristic approach used to obtain (\ref{E:qap}), one finds that an expansion of $D_a$, $E_a$ and their equations of motion in an inverse power series in $Z_0$ leads to equations of motion which are identical to equation (\ref{E:Ap}) for $A^{\prime}$ up to a different normalization of the source terms. Using equations (\ref{E:soldilaton}) and (\ref{E:qap}), one may easily read off the coefficients
\begin{align}
	Q_{D_a} &= \sqrt{2} Q_{A^{\prime}}, \\
	Q_{E_a} &= \left(\frac{\left(2 \alpha ^2-1\right)
		\left(\alpha ^2(v^2-1)+(v^2+2)\right)}{3 \left(\alpha ^2+1\right)} \right)
		Q_{A^{\prime}},
\end{align}
\end{subequations}
where terms of order $\mathcal{O}(Z_0^{-4})$ have been neglected. A more careful analysis following \cite{mydilaton} which leads to the same result has been left to appendix \ref{A:WKB}. There we also find that the regime of validity of equations (\ref{E:solutionnew}) is given by large $Z_0$ and $\alpha< Z_0^{2/3}$.

Putting together (\ref{E:constraints2}), (\ref{E:DbtoD1}), (\ref{E:EdtoE1})  and (\ref{E:solutionnew}) we find
\begin{align}
	Q_{A}&=\frac{1}{16}\pi \tilde{k} - \frac{i \alpha}{3 \tilde{k} z_0^2},\\
	Q_{D_1}&=Q_{A}+\frac{i\alpha}{4 \tilde{k} z_0^2},\\
	Q_{E_1}&=\frac{2}{3}\left((2+v^2)-(1-v^2)\alpha^2\right)Q_A+\frac{i \alpha}{\tilde{k} z_0^2}\left(\frac{1}{6}(1+5v^2)-\frac{1}{3}\alpha^2(1-v^2)\right).
\end{align}

As discussed earlier, we may now find the $z^4$ contribution $Q_{\mu\nu}$ to the metric fluctuations, $\hat{h}_{\mu\nu}$, due to the string by inverting the relations (\ref{E:defA}), (\ref{E:defD}) and (\ref{E:defE}). Plugging the resulting $Q_{\mu\nu}$ into (\ref{E:holographictmn}) one finds that the contribution of the quark motion to the Fourier transform of the energy momentum tensor
$\sqrt{\frac{1-v^2}{\lambda}}\langle \hat{T}_{quark\,\mu\nu}\rangle+\mathcal{O}\left(\left(\frac{T}{k}\right)^4\right)$ is given by
\begin{multline}
\label{E:emtensor}
		\frac{\tilde{k}}{24}
		\begin{pmatrix}
			-\left(2+v^2-(1-v^2)\frac{v^2 k_-^2}{\tilde{k}^2}\right) &
			3v-\frac{(1-v^2)v k_-^2}{\tilde{k}^2} &
			-\frac{(1-v^2) v k_{\bot} k_-}{\tilde{k}^2} &
			0
		\\
			3v-\frac{(1-v^2)v k_-^2}{\tilde{k}^2} &
			-(1+2v^2)+\frac{(1-v^2)k_-^2}{\tilde{k}^2} &
			\frac{(1-v^2)k_{\bot}{k_-}}{\tilde{k}^2} &
			0
		\\
			-\frac{(1-v^2) v k_{\bot} k_-}{\tilde{k}^2} &
			\frac{(1-v^2)k_{\bot}{k_-}}{\tilde{k}^2} &
			-\frac{(1-v^2)^2 k_-^2}{\tilde{k}^2} &
			0
		\\
			0 &
			0 &
			0 &
			-(1-v^2)
		\end{pmatrix}	
		\\+
		i v \pi k_- \left(\frac{T}{\tilde{k}}\right)^2
		\begin{pmatrix}
			\epsilon &
			\frac{1}{3}v-\frac{1}{9}(1-v^2)v\frac{k_-^2}{\tilde{k}^2} &
			\frac{k_{\bot}v}{2k_-}-\frac{k_- k_{\bot}(1-v^2) v}{9\tilde{k}^2} &
			0
		\\
			\frac{1}{3}v - \frac{1}{9}(1-v^2)v\frac{k_-^2}{\tilde{k}^2} &
			p_{k_-} &
			-\frac{k_{\bot}}{2 k_-}+\frac{1}{9}(1-v^2)\frac{k_- k_{\bot}}{\tilde{k}^2} &
			0
		\\
			\frac{k_{\bot}v}{2k_-}-\frac{k_- k_{\bot}(1-v^2) v}{9\tilde{k}^2} &
			-\frac{k_{\bot}}{2 k_-}+\frac{1}{9}(1-v^2)\frac{k_- k_{\bot}}{\tilde{k}^2} &
			p_{k_{\bot}} &
			0
		\\
			0 &
			0 &
			0 &
			p_{\theta}
		\end{pmatrix}
\end{multline}
where
\begin{align}
\epsilon &= \frac{1}{18}(5-11v^2)+\frac{1}{9}(1-v^2)v^2 \frac{k_-^2}{\tilde{k}^2},\\
p_{k_-} &= \frac{1}{18}(5v^2-11)+\frac{1}{9}(1-v^2)\frac{k_-^2}{\tilde{k}^2},\\
p_{k_{\bot}} &= \frac{1}{2}(1-v^2)-\frac{1}{9}(1-v^2)^2 \frac{k_-^2}{\tilde{k}^2},\\
p_\theta & = \frac{7}{18}(1-v^2).
\end{align}
It is satisfying to see that the leading $T=0$ contribution due to the quark motion is exactly what was predicted in \cite{Gubseremtensor} based on conformal invariance and Poincar\`{e} symmetry.
Fourier transforming the leading dissipative contribution to position space, we obtain equation (\ref{E:main1}).

\section{Acknowledgements}
I would like to thank O. Aharony, R. Brustein, A. Buchel, J. Friess, S. Gubser, G. Michalogiorgakis, S. Pufu and U. Wiedemann for useful correspondence and comments and M. Haack for many interesting discussions. I am supported in part by the German Science Foundation and by a Minerva fellowship.

\begin{appendix}
\section{\label{A:matrices}The equations of motion for the energy momentum tensor}
In order to decouple the components of equation (\ref{E:EOMmetric}), we have defined in equations (\ref{E:defA}) through (\ref{E:defE}) the variables $A$, $\vec{D}$ and $\vec{E}$. The resulting equations of motion were given in (\ref{E:EOMtensor}) and (\ref{E:constrainttensor}). The matrices $K_X$, $V_X$, $\tilde{K}_X$ and $\tilde{V}_X$ and the vectors $\vec{S}_X$ and $\vec{\tilde{S}}_X$ in these equations are given by
\begin{align}
	K_{A} &= 1 \,,\quad
	&V_{A} &= \frac{k^2}{g^2}\left(v^2 \left(\frac{k_-}{k}\right)^2-g\right) \,,\quad
	&S_A &= 1,\\
	K_{D} &=
		\begin{pmatrix}
			0 & 0\\
			0 & 1
		\end{pmatrix} \,,\quad
	&V_{D} &= \frac{k^2}{g^2}
		\begin{pmatrix}
			-g & g v^2 \left(\frac{k_1}{k}\right)^2 \\
			-1 & v^2 \left(\frac{k_1}{k}\right)^2
		\end{pmatrix} \,, \quad
	&\vec{S}_D &= \begin{pmatrix} 1 \\ 1 \end{pmatrix},
\end{align}
\begin{align}
	K_{E} &=
		\begin{pmatrix}
			\frac{3}{2} & 0 & 0 & 0\\
			0 & 0 & 0 & 0\\
			0 & 0 & \frac{1}{2} & 0\\
			0 & 0 & 0 & 1
		\end{pmatrix} \,, \quad
	\vec{S}_E = \begin{pmatrix}
		1+\frac{v^2}{g} \\ 1 \\ -1+v^2 - \frac{v^2}{g}
		\\ v^2 \left(-1+3\left(\frac{k_-}{k}\right)^2\right)
		\end{pmatrix},\\
	V_{E} &= \frac{k^2}{3 g^2}
		\begin{pmatrix}
			-2 g & 12 v^2 \left(\frac{k_-}{k}\right)^2 & 6 v^2 \left(\frac{k_-}{k}\right)^2 +2 g & 0 \\
			0 & 0 & 2 g & g\\
			0 & 0 & -2 g & -g \\
			2 g & -12 v^2 \left(\frac{k_-}{k}\right)^2 & 0 & 3 v^2 \left(\frac{k_-}{k}\right)^2+g
		\end{pmatrix},
\end{align}
\begin{align}
	\tilde{K}_{D} &= \begin{pmatrix}
		0 & 0 \\
		1 & -g
		\end{pmatrix} \,, \quad
	\vec{\tilde{S}}_{D}  =\begin{pmatrix}
		0 \\ \frac{z^2}{i v z_0^2 k_-}
		\end{pmatrix}, \quad
	\tilde{V}_D = 0,\\
	\tilde{K}_{E} &= \begin{pmatrix}
		0 & 1 & 1 & 0 \\
		-g & 0 & \quad -3 v^2 \left(\frac{k_-}{k}\right)^2 - g & \quad -g \\
		g & 0 & 2 & 0
		\end{pmatrix}, \,
	\vec{\tilde{S}}_{E} =
		\begin{pmatrix}
			\frac{i v z^2}{k_- z_0^2 g} \\
				\phantom{\Big|}
			-\frac{3 i v k_- z^2}{k^2 z_0^2}\left(\frac{v^2}{g}+1\right)\\
			z\left(1-\frac{v^2}{g}\right)\\
		\end{pmatrix},\\
	\tilde{V}_{E} &= \frac{1}{6 g}\begin{pmatrix}
		0 & -6 g' & -3 g' & 0\\
		-3 g g' & 18 v^2 \frac{k_-^2}{k^2} g' & 3 \left(3 v^2 \frac{k_-^2}{k^2}+g\right)g'
			& 0\\
		2 k^2 g z & \quad -12 k_-^2 v^2 z  & \quad -2 \left( 3 v^2 k_-^2 - g k^2\right) z & \,2 k^2 g z\\
		\end{pmatrix}.
\end{align}

\section{\label{A:WKB}WKB approximation}
In this section we would like to solve the equations of motion (\ref{E:Db}) and (\ref{E:Ed}) with boundary conditions such that $D_a(0)=E_a(0)=0$ and that $D_a$ and $E_a$ have only ingoing modes at the horizon. In effect, we are not interested in the full solutions to the equations of motion, but rather in the fourth order terms in a series expansion of $D_a$ and $E_a$ in small $Z$.

This problem has been addressed in \cite{mydilaton} for the related equation (\ref{E:Ap}). As discussed in the text, the latter equation of motion  has been solved by first obtaining the homogeneous solutions and then using them to construct the appropriate Greens function. Here we shall follow the same method, using some of the arguments made in \cite{mydilaton}.

First, we shall find the homogeneous solutions to equations (\ref{E:Db}) and (\ref{E:Ed}) by bringing them to Schr\"{o}dinger form, $\psi^{\prime\prime}+V\psi = 0$. This may be done by defining $\psi_X = \sqrt{\frac{b_X(Z)}{Z^3}}X$ with $X=E_d,D_b$, and
\begin{subequations}
\label{E:bs}
\begin{align}
	b_{D_a}(Z) &= g(Z)^{\alpha^2},\\
	b_{E_a}(Z) &= g(Z)^{-1-\frac{4}{3}\alpha\left(1+\alpha^2\right)},
\end{align}
\end{subequations}
and dividing the resulting equation by $\sqrt{\frac{Z^3}{b_X(Z)}}g(Z)$. The Schr\"{o}dinger equations obtained this way are characterized by the potentials
\begin{subequations}
\label{E:Vs}
\begin{align}
\label{E:VD}
	V_{D_b}(Z) & =V_0
		-\frac{4 \left(\alpha ^2+1\right)^2 \left(\frac{Z}{Z_0}\right)^6}{Z_0^2 g(Z)^2},\\
\label{E:VA}
	V_{E_d}(Z) & =V_0
		-\frac{64 \alpha ^4 \left(\alpha ^2+1\right)^2 \left(\frac{Z}{Z_0}\right) ^6}{9 Z_0^2 g(Z)^2},\\
\intertext{where}
\label{E:V0}
	V_0 &=-\frac{15}{4 Z^2}-\frac{1-(1+\alpha^2)\left(\frac{Z}{Z_0}\right)^4}{g(Z)^2}+\frac{4\left(\frac{Z}{Z_0}\right)^6}{Z_0^2 g(Z)^2}.
\end{align}
\end{subequations}
In equation (\ref{E:V0}), $V_{0}$ is the Schr\"odinger potential corresponding to the equation of motion for a massless scalar field in an AdS-SS background.

In the notation of equations (\ref{E:Vs}), it is clear that the $V_X$'s differ from $V_{0}$ only in their near horizon asymptotics. In \cite{mydilaton} it was shown that as long as the Schr\"odinger potential describing the asymptotically AdS system has a parametrically long flat region for which $V(Z) \simeq -1$ followed by a $V(Z)>0$ region near the horizon, then the $\mathcal{O}(Z^4)$ coefficient $Q_X$ in a small $Z$ expansion of $X$ is given by the $Z$ independent terms in
\begin{equation}
\label{E:formulaforQ}
	\frac{1}{16}\pi\tilde{k}^{-1}-\frac{1}{8}\int_{Z}^{Z_0^{2/3}} \frac{J_X(x) \sqrt{b_X(x)}}{x} K_2(x) dx,
\end{equation}
where exponentially suppressed terms, $e^{-Z_0^{2/3}}$, should be consistently neglected.
$J(x)$ is the non homogeneous term in the equation of motion for $X$. In our case (equations (\ref{E:Db}) and (\ref{E:Ed})) we find
\begin{align}
	J_{D_a}(Z) &= \sqrt{2} J_0(Z),\\
	J_{E_a}(Z) &= \left(
	\frac{\left(2 \alpha ^2-1\right) \left(\left(\alpha ^2+1\right) v^2-\alpha ^2+2\right)}{3 \left(\alpha ^2+1\right)}
	-
	\frac{2 v^2 Z^4 \left(\alpha ^2+1\right) \left(2 \alpha ^2-1\right)}{3 \left(Z^4-Z_0^4\right)}
	\right) J_0(Z),\\
\intertext{where}
	J_{0}(Z)&=\frac{Z}{\tilde{k}}e^{-i k_- \xi(Z/\tilde{k})}{g(Z)}
\end{align}
is the source term for the equation of motion for the $A'$ component of the metric (which coincides with the equation of motion of the dilaton). $b(x)$ is given in equations (\ref{E:bs}) and $K_2(x)$ is a modified Bessel function of the second kind.

Clearly, as long as $\alpha < Z_0^{2/3}$ the potentials (\ref{E:VD}) and (\ref{E:VA}) satisfy the flatness and positivity requirements discussed above. Therefore, one may use formula (\ref{E:formulaforQ}) to obtain the fourth order coefficients of $D_a$, $E_a$ and $A^{\prime}$. The $A^{\prime}$ case has been discussed in \cite{mydilaton}, where it was shown that as long as $Z_0<\alpha^2$ one may expand the integral in (\ref{E:formulaforQ}) in a power series in $Z_0$ owing to the exponential decay of the modified Bessel function $K_2(x)$. One finds $\tilde{k} Q_{A^{\prime}} = \frac{1}{16}\pi -\frac{i \alpha}{3 Z_0^{2}}+ \mathcal{O}(Z_0^{-4})$. To obtain the other two solutions, we note that up to order $\mathcal{O}(Z_0^{-4})$, the expressions $J_{D_a} \sqrt{b_{D_a}}$ and $J_{E_a} \sqrt{b_{E_a}}$ differ from $J_{A^{\prime}} \sqrt{b_{A^{\prime}}} = J_0 \sqrt{g(Z)}$ by a constant multiplicative term. This leads us to equations (\ref{E:solutionnew}).

\end{appendix}

\bibliography{JQ}

\end{document}